\documentclass{article}

\usepackage{arxiv}

\usepackage[utf8]{inputenc} 
\usepackage[T1]{fontenc}    
\usepackage{hyperref}       
\usepackage{url}            
\usepackage{booktabs}       
\usepackage{amsfonts}       
\usepackage{nicefrac}       
\usepackage{microtype}      
\usepackage{lipsum}
\usepackage{graphicx}
\usepackage{amsmath}     
\usepackage{amssymb}     
\usepackage{algorithm}
\usepackage{algpseudocode}
\usepackage{subfigure}
\usepackage{multirow}
\graphicspath{ {./images/} }

\title{Take Package as Language: Anomaly Detection Using Transformer}

\author{
 Jie Huang \\
  Chengdu University of Information Technology\\
}

\begin{document}
\maketitle
\begin{abstract}
Network data packet anomaly detection faces numerous challenges, including exploring new anomaly supervision signals, researching weakly supervised anomaly detection, and improving model interpretability. This paper proposes NIDS-GPT, a GPT-based causal language model for network intrusion detection. Unlike previous work, NIDS-GPT innovatively treats each number in the packet as an independent "word" rather than packet fields, enabling a more fine-grained data representation. We adopt an improved GPT-2 model and design special tokenizers and embedding layers to better capture the structure and semantics of network data. NIDS-GPT has good scalability, supports unsupervised pre-training, and enhances model interpretability through attention weight visualization. Experiments on the CICIDS2017 and car-hacking datasets show that NIDS-GPT achieves 100\% accuracy under extreme imbalance conditions, far surpassing traditional methods; it also achieves over 90\% accuracy in one-shot learning. These results demonstrate NIDS-GPT's excellent performance and potential in handling complex network anomaly detection tasks, especially in data-imbalanced and resource-constrained scenarios. The code is available at \url{https://github.com/woshixiaobai2019/nids-gpt.git}
\end{abstract}


\section{Introduction}
Intrusion Detection Systems (IDS) play a crucial role in network security defense, serving as a key component in identifying and preventing various network threats~\cite{abdulganiyu2023systematic}. Among these, Network Intrusion Detection Systems (NIDS) focus on monitoring network traffic, analyzing suspicious activities, and promptly detecting potential security vulnerabilities and attack behaviors~\cite{verma2022inids}. Essentially, NIDS performs a classification task, determining the category of data packets, such as normal packets, DOS packets, or DDOS packets.

In the training and evaluation process of NIDS, experts typically rely on simple annotations of network packets, categorizing them as normal or abnormal, or further classifying them into specific anomaly types~\cite{sharafaldin2018toward,song2020vehicle}. However, this coarse-grained annotation method overlooks the rich information and complex associations within packets, such as interdependencies between packet fields, temporal relationships, and semantic connections. This results in relatively weak supervision signals, failing to fully exploit the inherent features and patterns of packets, thus limiting the detection performance and generalization ability of NIDS~\cite{pang2021deep}. It also makes it difficult for NIDS to learn the importance and contribution of various packet fields, hindering effective identification of anomalous changes and combinations in key fields, thereby reducing the ability to detect unknown attacks. Furthermore, NIDS faces the challenge of extreme data imbalance~\cite{pang2021deep,gupta2022cse}. In real network environments, the proportion of attack or anomalous packets is usually very small, and there may even be zero-day attacks or unknown threats. This data imbalance poses significant challenges to the training and optimization of NIDS, making it difficult to learn comprehensive and robust detection models, potentially resulting in high false negative and false positive rates when facing new types of attacks or variants.

To address these challenges, numerous methods have been proposed. To tackle the weak supervision signal problem, some studies have employed feature engineering techniques such as dimensionality reduction~\cite{waskle2020intrusion} and feature selection~\cite{sharma2020optimization} to simplify data complexity and facilitate model learning of intrinsic packet patterns. On the other hand, to address the data imbalance challenge, researchers have explored various data balancing and data augmentation techniques. Generative Adversarial Networks (GAN), Variational Autoencoders (VAE), and Gaussian Mixture Models (GMM) have been used to generate synthetic minority class samples to balance datasets~\cite{liu2022intrusion,lee2021gan,wang2023intrusion}. Although these methods have made some progress, they still rely heavily on large amounts of labeled data and remain ineffective under extreme data imbalance conditions (e.g., 1000:1).

In recent years, large-scale pre-trained language models have made breakthrough progress in the field of natural language processing. In 2018, OpenAI's GPT (Generative Pre-trained Transformer) model pioneered a new paradigm of enhancing model performance through large-scale unsupervised pre-training~\cite{radford2018improving}. In the same year, Google's BERT (Bidirectional Encoder Representations from Transformers) model further improved the pre-training strategy by introducing bidirectional context encoding, significantly enhancing performance on multiple NLP tasks~\cite{47751}. The success of these models inspired the development of even larger-scale language models. Subsequently, larger models like GPT-2 and GPT-3 emerged, demonstrating astonishing language understanding and generation capabilities~\cite{brown2020language}. Recently, Large Language Models (LLMs) represented by ChatGPT have garnered widespread attention. These models, trained on massive text data, have opened new directions for artificial intelligence development~\cite{wei2022emergent}.

\begin{figure}[t]  
    \centering  
    \includegraphics[width=0.97\textwidth]{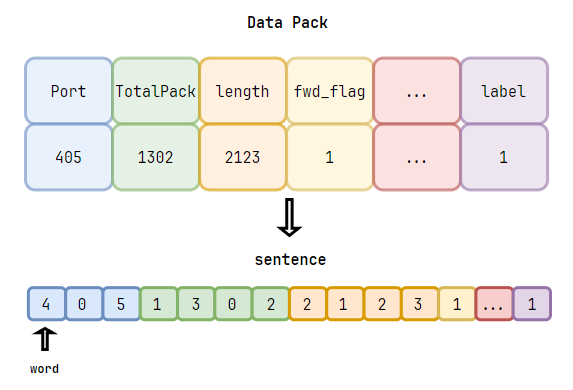}  
    \caption{The packet language}  
    \label{fig:packet_language}  
\end{figure}

\begin{figure}[h!]
  \centering
  \subfigure[Direct classification-based method]{
    \label{fig:motivation:direct_cls}
    \includegraphics[width=0.97\linewidth]{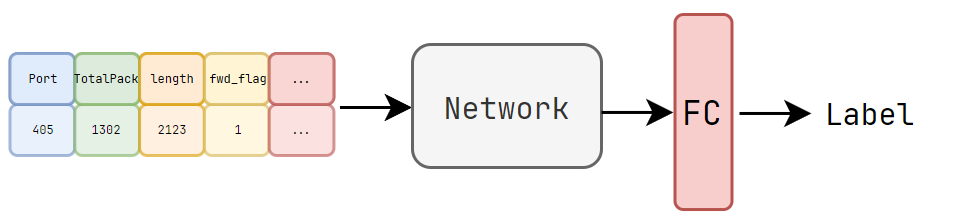}
  }
  \subfigure[BERT-based method]{
    \label{fig:motivation:bert}
    \includegraphics[width=0.97\linewidth]{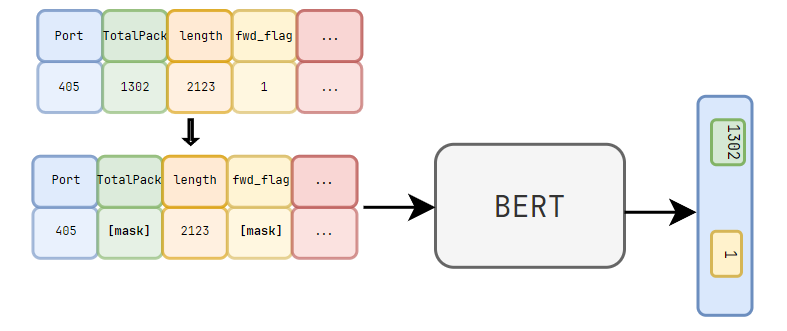}
  }
  \subfigure[Our proposed NIDS-GPT]{
    \label{fig:motivation:nids-gpt}
    \includegraphics[width=0.9\linewidth]{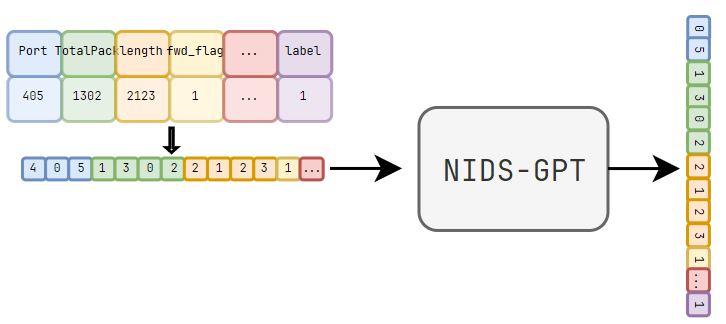}
  }
  \caption{
    Comparison of our method with existing methods
  }
  \label{fig:motivation}
\end{figure}

Inspired by these studies, we propose a language modeling-based NIDS model called NIDS-GPT. Specifically, we treat each number in the data packet as a "word". For example, we consider the port "22" as two words, "2" and "2", and the combination of all words in each field of the packet forms a "sentence" with rich semantics, which we liken to the language of communication between computers. As shown in Figure \ref{fig:packet_language}, a data packet is represented as a "sentence" composed of numeric strings. Based on this idea, we utilize language models to learn the rich semantics of these "sentences".

Our proposed NIDS-GPT model includes a packet tokenization layer, a packet embedding layer, and a classic causal language model layer. The tokenization layer can mark field boundaries and implicitly encode field order information, providing additional structured information to the model. The embedding layer uses a multi-faceted embedding strategy, including word embeddings, numerical position embeddings, and field position embeddings. This multi-dimensional representation enables the model to comprehensively understand the complexity of network data. In addition to model innovation, inspired by language modeling, we set the objective function to predict the next "word". We append the packet category to the end of the packet as the last "word" of the packet. We optimize our model by calculating the negative log-likelihood loss for all "words", not just the label "word". Figure \ref{fig:motivation} illustrates the differences between our method and existing methods. To comprehensively evaluate the model's learning ability, we designed a series of tasks, including imbalanced data learning, one-shot learning, and transfer learning. Our experimental results demonstrate that the NIDS-GPT model possesses powerful learning capabilities.

The main contributions of this paper are as follows:
\begin{itemize}
\item We propose a GPT-based NIDS model (NIDS-GPT) for network packet anomaly detection.
\item We introduce a new training objective.
\item We design a series of tasks to comprehensively evaluate the performance of the NIDS-GPT model in imbalanced data learning, one-shot learning, and transfer learning.
\item We conduct preliminary interpretability exploration on the trained model.
\end{itemize}

\section{Related Work}

This section will introduce existing methods for network intrusion detection, divided into four parts: traditional deep learning models (such as CNN, LSTM, and RNN), Transformer-based language models (such as GPT and BERT), methods based on modern large language models (LLMs), and the NIDS-GPT model proposed in this paper.

\textbf{Traditional Deep Learning-based Methods:}
The literature ~\cite{yang2022transfer} proposed a CNN-based NIDS detection method that first converts network traffic data into images and then uses traditional CNN networks for learning, achieving good results. However, this method requires image conversion processing and cannot directly learn from raw data. The literature ~\cite{kim2020cnn} proposed a CNN-based deep learning intrusion detection model focusing on Denial of Service (DoS) attacks, designing 18 different experimental scenarios and testing binary and multi-class classification tasks. However, this model only focuses on DoS attacks and has not tested other types of network attacks. The literature ~\cite{hnamte2023novel} proposed a novel two-stage deep learning model combining Long Short-Term Memory (LSTM) networks with Autoencoders (AE), using data filtering to prevent overfitting and underfitting, achieving an effective balance between dimensionality reduction and feature retention on highly imbalanced datasets. The literature ~\cite{sivamohan2021effective} proposed a recurrent neural network based on Bidirectional Long Short-Term Memory (BiLSTM) networks, using Random Forest and Principal Component Analysis (PCA) algorithms for feature selection on the CICIDS2017 dataset. Although these papers have shown some improvement in accuracy, they essentially have not solved the problem of sparse supervision signals.

\textbf{BERT-based Methods:}
The literature ~\cite{nwafor2022canbert} proposed a model called CanBERT based on BERT for Controller Area Network (CAN) intrusion detection, as shown in Figure \ref{fig:motivation:bert}. It uses Masked Language Modeling (MLM) pre-training method, randomly masking certain fields in the packet and then predicting these fields. It also adopted some advanced data processing techniques such as BPE tokenization ~\cite{sennrich2015neural} and RoPE ~\cite{su2024roformer} positional encoding. They trained a BERT model from scratch, fine-tuning the top layer of the model after pre-training to achieve classification tasks, obtaining good results. The literature ~\cite{alkhatib2022can} also proposed a similar approach but trained for a simpler binary classification fine-tuning task. The paper compared with a large number of other traditional methods, further demonstrating the potential of using language models to learn traffic data.

\textbf{GPT-based Methods:}
There are also some works similar to ours, for example, the literature ~\cite{nam2021intrusion} proposed an intrusion detection method based on bidirectional Generative Pre-trained Transformers (GPT), using two interconnected GPT networks to form a bidirectional structure to more accurately predict each CAN identifier (ID), thereby being able to stably estimate the entire CAN ID sequence. This method identifies potential attacks by calculating the negative log-likelihood (NLL) value of the CAN ID sequence and comparing it with a preset threshold. Experimental results show that compared to unidirectional GPT networks, bidirectional GPT networks demonstrate significant advantages in detection performance, especially when dealing with situations containing very few attack signals. Furthermore, experiments also show that as the amount of training data increases, the F-measure performance of this method also improves, further proving the effectiveness of language modeling on large-scale datasets. However, unfortunately, the threshold-based judgment can only be used for binary classification tasks and strongly relies on empirical values. The paper also did not use supervised fine-tuning and did not provide the tokenization method used, lacking many experimental details. As for the method itself, its model structure and training objectives are also different from our work.

\textbf{Large Language Model-based Methods:}
In recent years, with the rise of large language models like ChatGPT, work on using large language models for anomaly detection has gradually emerged. The literature ~\cite{shi2024language} proposed a framework called LAMP that utilizes the reasoning ability of large language models to improve event sequence prediction. Specifically, the event model first proposes predictions for future events based on past events, then the language model, guided by learning from a few expert-annotated examples, provides possible reasons for each prediction; next, the search module finds historical events that match these reasons; finally, the scoring function evaluates whether these historical events can actually lead to the occurrence of predicted future events. The literature ~\cite{jin2023time} proposed a method called TIME-LLM for general time series prediction by reprogramming large language models (LLMs). TIME-LLM aims to leverage LLMs' powerful pattern recognition and reasoning capabilities on complex sequences by converting time series data into text prototypes as input to frozen LLMs to solve time series prediction problems for different tasks and applications. Although they have achieved certain effects, the current training and inference costs of large language models are too high, which is a very realistic problem.

\section{Method}

\begin{figure}[t]
    \centering
    \includegraphics[width=1\textwidth]{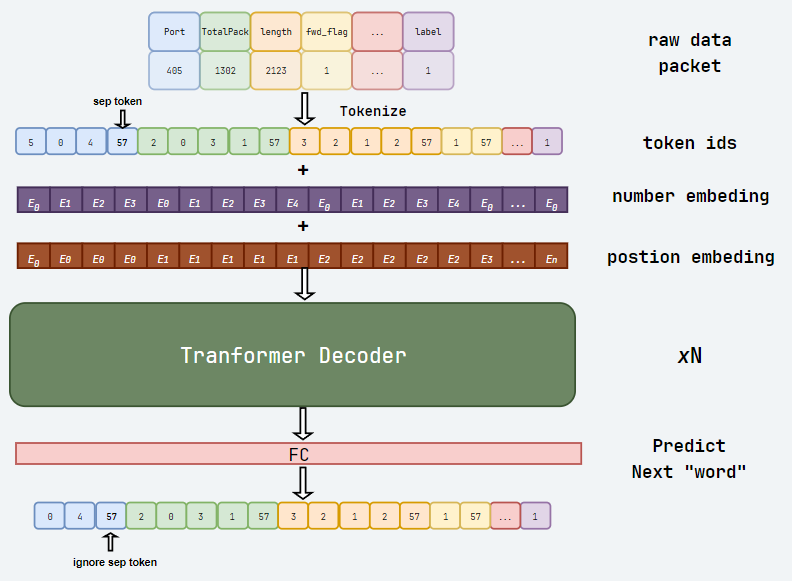}
    \caption{Overall model design}
    \label{fig:model-architecture}
\end{figure}

This section will provide a detailed introduction to our proposed NIDS-GPT model. We will first give an overall overview, then elaborate on our core design - the tokenizer and embedding layer.

\subsection{Overview}
For a network packet $X=\{x_1,x_2,x_3,\ldots,x_n\}$ with N fields, each field $x_i$ can be represented by a positive integer (non-numeric types can be converted to numbers). Existing methods use direct classification or BERT's field-level training approach ~\cite{yang2022transfer,nwafor2022canbert,alkhatib2022can}. Our method first converts the original packet $D$ into a series of tokens, a process that can be represented by the function $f_{tokenize}$, whose output is a sequence of token IDs. Unlike methods using BPE tokenization and RoPE positional encoding ~\cite{nwafor2022canbert}, we implemented our own tokenizer and positional encoding method. We generate two types of embedding representations for each token ID: numeric embedding $E_n$ and positional embedding $E_p$. The numeric embedding $E_n$ captures the numerical information of the token, while the positional embedding $E_p$ encodes the position information of the token in the sequence. These two embeddings are combined through addition to form the final embedding representation $E = E_n + E_p$. The resulting embedding sequence is then input into a multi-layer Transformer decoder. The Transformer decoder processes these embeddings through its self-attention mechanism, effectively capturing long-distance dependencies and complex patterns in the packet. The output of the Transformer decoder goes through a fully connected layer (FC) to predict the next "word" (i.e., the next token ID) in the sequence. This prediction process can be formalized as a conditional probability distribution $P(t_{i+1}|t_1, ..., t_i)$, where $t_i$ represents the token ID at the $i$-th position. During the training phase, the model optimizes by minimizing the cross-entropy loss of each "word" prediction. It's worth mentioning that we treat the packet's label as the last "word" of the packet, which nicely implements the classification task. We show the entire processing flow in Figure \ref{fig:model-architecture}.

\subsection{Network Packet Tokenizer}

Our method first views the network packet as a collection of a series of fields. Given an input sequence $X = (x_1, x_2, \ldots, x_n)$, where $x_i$ represents different fields of the network packet (such as source port, destination port, IP address, etc.). These fields usually appear in numerical form, but we choose to view them as strings for more fine-grained processing. For example, the port number 406 is viewed as the string "406" rather than a single integer value.

To better preserve the numerical features of the data in subsequent positional encoding, we introduce an innovative field reversal mechanism. For each field $x_i$, we reverse its character order:

\begin{equation}
    x_i' = reverse(x_i)
\end{equation}

where $reverse()$ represents the string reversal operation. The advantage of this approach is that regardless of how many digits the original value has, the most significant digit (ones place) always appears in a fixed position. This consistency helps the model better understand and learn the structure of the numbers.

After field reversal, we begin to construct the tokenized sequence $R$. For each field except the last one, we convert each of its characters to the corresponding integer value and add them to the result sequence in order (the $token ids$ in Figure \ref{fig:model-architecture}):

\begin{equation}
    R = R \cup \{int(c) | c \in x_i'\} \cup \{S_i\}, \quad \forall i \in [1, n-1]
\end{equation}

where $int()$ converts characters to integers, and $S_i$ is a special separator token. This separator token is not static but dynamically changing:

\begin{equation}
    S_i = S_0 + i - 1
\end{equation}

where $S_0$ is the initial separator token value. This design of dynamic separator tokens not only marks the boundaries of fields but also implicitly encodes the order information of fields, providing additional structured information for the model.

For the last field $x_n$ in the packet, we adopt special processing. Unlike other fields, the last field is directly added to the end of the sequence as an integer, without reversal operation:

\begin{equation}
    R = R \cup \{int(x_n)\}
\end{equation}

This processing method conveniently transforms the label classification into the task of predicting the next "word". Finally, considering that deep learning models usually require fixed-length inputs, we introduce a padding mechanism. If the constructed sequence length is less than the predefined length $L$, we add special padding tokens at the end of the sequence.

\subsection{Network Packet Embedding}

Before inputting the tokenized sequence into the deep learning model, we need to convert these discrete tokens into continuous vector representations. At the same time, to enable the model to understand the structure of the input sequence, we also need to introduce positional information. This section will detail our embedding and positional encoding methods, which are specifically designed to handle the special structure and characteristics of network packets.

A key feature of network packets is that they contain multiple different types of fields, each of which may contain multiple digits. Traditional embedding methods often struggle to capture both this hierarchical structure and the internal structure of the numbers simultaneously. To address this issue, we propose a multi-faceted embedding strategy, including word embeddings, numeric position embeddings, and field position embeddings. These three types of embeddings work together to provide the model with rich feature representations.

First, for the input sequence $X = (x_1, x_2, \ldots, x_n)$, we use word embeddings to map each token to a $d$-dimensional vector space:

\begin{equation}
    E_w(x_i) = W_e \cdot x_i
\end{equation}

where $W_e \in \mathbb{R}^{V \times d}$ is the word embedding matrix, $V$ is the vocabulary size, and $d$ is the embedding dimension. The main role of word embeddings is to capture the semantic information of each token. In network data analysis, this can help the model understand the meaning of different numbers and special tokens (such as separators).

Next, we introduce numeric position embeddings. This is an innovation in our method, aimed at capturing the position of each digit in its original field (ones place, tens place, hundreds place, etc.). This design is based on the observation that in many network protocols, different positions of a number may have different semantic importance. For example, in an IP address, the highest and lowest bits may carry different network topology information. For each token $x_i$, we have a corresponding numeric position index $p_i$:

\begin{equation}
    E_n(p_i) = N_e \cdot p_i
\end{equation}

where $N_e \in \mathbb{R}^{M \times d}$ is the numeric position embedding matrix, and $M$ is the maximum numeric length. In this way, the model can distinguish the importance of the same number at different positions, such as distinguishing whether the "4" in "406" is in the hundreds place or the ones place.

Finally, we use field position embeddings to encode which network packet field (such as port number, IP address, etc.) each token belongs to. This embedding allows the model to understand the overall structure of the packet and potentially learn relationships between different fields. For example, there may be certain patterns between source IP addresses and destination IP addresses. For each token $x_i$, we have a corresponding field position index $f_i$:

\begin{equation}
    E_p(f_i) = F_e \cdot f_i
\end{equation}

where $F_e \in \mathbb{R}^{L \times d}$ is the field position embedding matrix, and $L$ is the maximum sequence length. Field position embeddings allow the model to distinguish tokens from different fields, even if these tokens may have the same numerical value. For example, it can help the model distinguish between the same number in the source port and destination port.

The combination of these three types of embeddings provides a comprehensive representation for each token. The final token representation is the sum of these three embeddings:

\begin{equation}
    E(x_i, p_i, f_i) = E_w(x_i) + E_n(p_i) + E_p(f_i)
\end{equation}

It's worth noting that the numeric position embedding $E_n(p_i)$ is set to a zero vector when processing padding tokens or non-numeric fields (such as protocol type) to avoid introducing irrelevant position information. This design ensures that the model only considers numeric position information when processing numeric data.

\subsection{Objective Function and Inference}

Our model adopts an objective function similar to GPT during training, while focusing on label prediction during the inference phase. The training process aims to maximize the likelihood of predicting the next word in the sequence:
\begin{equation}
\mathcal{L} = \max_\theta \sum_{t=1}^{T} \log P(x_t|x_{<t}; \theta)
\end{equation}
where $x_t$ is the $t$-th word in the sequence, $x_{<t}$ represents all words before $t$, and $\theta$ are the model parameters. During the inference phase, our model focuses solely on predicting the label field. Assuming the label field is at position $l$, we define the inference process as:
\begin{equation}
\hat{y}_{\text{label}} = {argmax}_{y \in \mathcal{Y}} P(y|x_1, x_2, ..., x_T; \theta) = {argmax}_{y \in \mathcal{Y}} p_l
\end{equation}
where $\hat{y}_{\text{label}}$ is the predicted label, $\mathcal{Y}$ is the set of all possible labels, and $p_l$ is the probability distribution at the label position.

\subsection{Evaluation Metrics}
\label{subsec:evaluation-metrics}

To comprehensively evaluate our model's performance, we adopt the following three main metrics: Precision, Recall, and F1 score. These metrics are suitable for our classification task, which is predicting the next character in CAN data packets.

\paragraph{Precision}
Precision measures the proportion of actual positive samples among those predicted as positive by the model. The formal definition is as follows:

\begin{equation}
    \text{Precision} = \frac{TP}{TP + FP}
\end{equation}

where TP (True Positive) represents the number of samples correctly predicted as positive, and FP (False Positive) represents the number of samples incorrectly predicted as positive.

\paragraph{Recall}
Recall measures the proportion of samples correctly predicted by the model among those that are actually positive. The formal definition is as follows:

\begin{equation}
    \text{Recall} = \frac{TP}{TP + FN}
\end{equation}

where FN (False Negative) represents the number of samples incorrectly predicted as negative.

\paragraph{F1 Score}
The F1 score is the harmonic mean of precision and recall, providing a comprehensive performance measure. The formal definition is as follows:

\begin{equation}
    \text{F1} = 2 \cdot \frac{\text{Precision} \cdot \text{Recall}}{\text{Precision} + \text{Recall}}
\end{equation}

For multi-class classification problems, we calculate the precision, recall, and F1 score for each category, and then take the weighted average to obtain the overall performance metrics. Assuming there are $K$ categories, with $n_k$ samples in the $k$-th category, and a total of $N = \sum_{k=1}^K n_k$ samples, the weighted average F1 score is calculated as follows:

\begin{equation}
    \text{Macro-F1} = \sum_{k=1}^K \frac{n_k}{N} \cdot \text{F1}_k
\end{equation}

where $\text{F1}_k$ is the F1 score of the $k$-th category.

Through these evaluation metrics, we can comprehensively assess the model's performance in the CAN data packet character prediction task, including the accuracy and completeness of predictions. These metrics can help us compare the performance of different models or different configurations and guide the improvement and optimization of the model.

\section{Experiments}

\begin{figure}[h]
    \centering
    \includegraphics[width=0.5\textwidth]{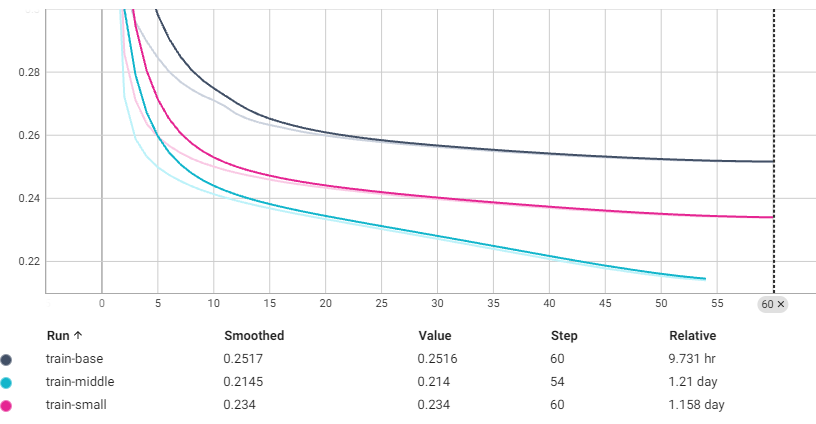}
    \caption{Training log}
    \label{fig:train-log}
\end{figure}

\begin{table}[h!]
\centering
\caption{Model configurations}
\begin{tabular}{ccccc}
\hline
Model size & n\_layers & n\_heads & emb\_size & seq\_len \\
\hline
base   & 6  & 8  & 128 & 256 \\
small  & 8  & 16 & 256 & 256 \\
middle & 10 & 32 & 512 & 256 \\
\hline
\end{tabular}
\label{table:model_size}
\end{table}

\begin{table}[h!]
\centering
\caption{Hyperparameter settings}
\begin{tabular}{cccccc}
\hline
epochs & learning\_rate & optimizer & learning\_decay & batch\_size & MLP ratio \\
60 & 1e-4(3e-5 for middle) & adam & cosine & 128 & $\frac{8}{3}$ \\
\hline
\end{tabular}
\label{table:train_setting}
\end{table}

\begin{table}[htbp]
\centering
\caption{Dataset split details for different tasks}
\begin{tabular}{lcccccccc}
\hline
\multirow{2}{*}{Category} & \multicolumn{4}{c}{Training} & \multicolumn{4}{c}{Testing} \\
\cline{2-9}
 & 0.001 & 0.0005 & 0.0002 & one-shot & 0.001 & 0.0005 & 0.0002 & one-shot \\
\hline
BENIGN & 18402 & 80000 & 79999 & 5000 & 20000 & 20000 & 20000 & 20000 \\
DoS slowloris & 27 & 29 & 17 & 1 & 20 & 7 & 4 & 4 \\
DoS Slowhttptest & 19 & 29 & 17 & 1 & 20 & 7 & 4 & 4 \\
PortScan & 13 & 29 & 17 & 1 & 20 & 7 & 4 & 4 \\
Bot & 22 & 28 & 17 & 1 & 20 & 8 & 4 & 4 \\
DoS Hulk & 19 & 29 & 17 & 1 & 20 & 7 & 4 & 4 \\
DoS GoldenEye & 22 & 29 & 17  & 1 & 20 & 7 & 4 & 4 \\
Web Attack Brute Force  & 17 & 29 & 17  & 1 & 20 & 7 & 4 & 4 \\
Web Attack Sql Injection  & 0 & 0 & 17 & 1 & 0 & 0 & 4 & 4 \\
Infiltration & 0 & 28 & 17 & 1 & 0 & 8 & 4 & 4 \\
Web Attack XSS & 16 & 29 & 16 & 1 & 20 & 7 & 5 & 5 \\
DDoS   & 12 & 29 & 16 & 1 & 20 & 7 & 5 & 5 \\
\hline
\end{tabular}
\label{tab:data_detail}
\end{table}

In this chapter, we will provide a detailed introduction to our experiments. We conducted extensive experiments on two standard datasets, and the results demonstrate the effectiveness of our method.

\textbf{Experimental Environment:} All experiments in this section are based on Python 3.8, using standard data processing libraries such as numpy, pandas, and the PyTorch deep learning framework.

\textbf{Model Configuration:} As shown in Table \ref{table:model_size}, we trained models of three different sizes, gradually increasing the width and depth of our model. Here, n\_layers represents the number of transformer decoder layers, n\_heads represents the number of multi-head attention heads, emb\_size is the dimension of the word embedding layer, and seq\_len is the maximum input data length.

\textbf{Hyperparameter Configuration:} As shown in Table \ref{table:train_setting}, we chose the Adam optimizer. Notably, for larger models, we used a smaller learning rate and employed linear warm-up with cosine decay. Since we used the SwiGLU~\cite{shazeer2020glu} activation function in the fully connected layer, we set a universal MLP ratio value.

\subsection{Network Traffic Packet Detection Experiment}

\textbf{Dataset:} This experiment used the CIC-IDS2017 dataset provided by the Canadian Institute for Cybersecurity. This dataset aims to provide reliable benchmark data for research on Intrusion Detection Systems (IDS) and Intrusion Prevention Systems (IPS). Compared to previous intrusion detection datasets, CIC-IDS2017 covers a variety of latest attack types, including brute force attacks (FTP and SSH), Denial of Service (DoS), Distributed Denial of Service (DDoS), Web attacks, infiltration attacks, botnets, etc., while also including naturally generated normal background traffic. The dataset is modeled based on real user behavior, capturing network traffic over 5 days (July 3-7, 2017), extracting over 80 network flow features using the CICFlowMeter tool, and providing detailed labels and metadata. Moreover, the CIC-IDS2017 dataset meets the needs of modern intrusion detection system evaluation in terms of network configuration, traffic capture, protocol coverage, and attack diversity, effectively supporting the development and validation of anomaly detection-based intrusion detection algorithms.

\textbf{Experimental Setup:} This experiment mainly follows the approach in the literature ~\cite{ferdigg2022self}, verifying the model's learning effect by setting imbalanced datasets. In ~\cite{ferdigg2022self}, the extreme imbalance ratio is 0.01, while we further increased the experimental difficulty by setting three imbalance ratios: 0.001, 0.0005, and 0.0002. Additionally, to verify the effect in the most extreme case, we also conducted a one-shot experiment, using only one sample for model learning. To more accurately assess the model's ability in single-sample learning situations, we conducted 10 random sampling experiments and took the average. Unlike ~\cite{ferdigg2022self}, we adopted multi-classification task metrics, which are more consistent with real environments. The final partitioned dataset is shown in Table \ref{tab:data_detail}.

\textbf{Training Log:} The training log of the entire model is shown in Figure \ref{fig:train-log}. It can be seen that the largest model (middle) has the lowest negative log-likelihood (NLL) on the training set, demonstrating the scalability of our model. It's worth mentioning that the experimental results we present subsequently are only using our smallest model (base).

\textbf{Results Analysis:} We compared our method with several existing methods under different imbalance ratios, including ET, RandomForest, optimized\_RF~\cite{goryunov2020synthesis}, and CNN\_BiLSTM~\cite{get2023deep}. The experimental results are shown in Tables \ref{table:result-cicids-1}, \ref{table:result-cicids-5}, \ref{table:result-cicids-2}, and \ref{table:result-cicids-oneshot}. From the tables, it can be seen that under imbalance ratios of 0.001, 0.0005, and 0.0002, our method achieved 1.00 in Precision, Recall, and F1-score, far surpassing other methods. This indicates that our method can effectively capture minority class samples in extremely imbalanced datasets while maintaining high overall classification performance. In contrast, existing methods such as ET and RandomForest, although performing reasonably well in Precision, showed certain deficiencies in Recall and F1-score, especially when the data imbalance intensified (imbalance ratio of 0.0002), their performance declined significantly. The optimized\_RF and CNN\_BiLSTM methods performed even less satisfactorily in various metrics, particularly in Recall and F1-score, struggling to effectively handle highly imbalanced datasets. Furthermore, the results of the one-shot experiment (Table \ref{table:result-cicids-oneshot}) further verify the robustness of our method. With only one negative sample for learning, our method still significantly outperformed other methods, achieving excellent scores of 0.91 in Precision, 0.82 in Recall, and 0.84 in F1-score. The performance of other methods was generally lower, especially optimized\_RF and CNN\_BiLSTM, which performed almost identically in the one-shot experiment, with Precision and Recall of only 0.08, and F1-score of only 0.08. In summary, our method demonstrated excellent performance under various experimental conditions, especially in handling highly imbalanced datasets, maintaining high classification accuracy and recall rates, significantly outperforming several existing mainstream methods.

\begin{table}
\centering
\caption{Comparison with other methods - Imbalance ratio: 0.001}
\begin{tabular}{ccccc}
\hline
method & avg & Precision & Recall & F1-score \\
\hline
ET & macro avg & 0.84 & 0.78 & 0.81 \\
RandomForest & macro avg & 0.83 & 0.76 & 0.79 \\
optimized\_RF~\cite{goryunov2020synthesis} & macro avg & 0.84 & 0.63 & 0.71 \\
CNN\_BiLSTM~\cite{get2023deep} & macro avg & 0.48 & 0.31 & 0.35 \\
\textbf{Ours} & macro avg & \textbf{1.00} & \textbf{1.00} & \textbf{1.00} \\
\hline
\end{tabular}
\label{table:result-cicids-1}
\end{table}

\begin{table}
\centering
\caption{Comparison with other methods - Imbalance ratio: 0.0005}
\begin{tabular}{ccccc}
\hline
method & avg & Precision & Recall & F1-score \\
\hline
ET & macro avg & 0.86 & 0.73 & 0.78 \\
RandomForest & macro avg & 0.94 & 0.74 & 0.90 \\
optimized\_RF~\cite{goryunov2020synthesis} & macro avg & 0.94 & 0.71 & 0.79 \\
CNN\_BiLSTM~\cite{get2023deep} & macro avg & 0.53 & 0.42 & 0.45 \\
\textbf{Ours} & macro avg & \textbf{1.00} & \textbf{1.00} & \textbf{1.00} \\
\hline
\end{tabular}
\label{table:result-cicids-5}
\end{table}

\begin{table}
\centering
\caption{Comparison with other methods - Imbalance ratio: 0.0002}
\begin{tabular}{ccccc}
\hline
method & avg & Precision & Recall & F1-score \\
\hline
ET & macro avg & 0.87 & 0.70 & 0.76 \\
RandomForest & macro avg & 0.87 & 0.62 & 0.76 \\
optimized\_RF~\cite{goryunov2020synthesis} & macro avg & 0.76 & 0.54 & 0.60 \\
CNN\_BiLSTM~\cite{get2023deep} & macro avg & 0.09 & 0.12 & 0.09 \\
\textbf{Ours} & macro avg & \textbf{1.00} & \textbf{1.00} & \textbf{1.00} \\
\hline
\end{tabular}
\label{table:result-cicids-2}
\end{table}

\begin{table}
\centering
\caption{Comparison with other methods - One-shot (average of 10 times)}
\begin{tabular}{ccccc}
\hline
method & Precision & Recall & F1-score \\
\hline
ET & 0.23 & 0.31 & 0.27 \\
RandomForest & 0.33 & 0.26 & 0.34 \\
optimized\_RF~\cite{goryunov2020synthesis} & 0.08 & 0.08 & 0.08 \\
CNN\_BiLSTM~\cite{get2023deep} & 0.08 & 0.08 & 0.08 \\
\textbf{Ours} & \textbf{0.91} & \textbf{0.82} & \textbf{0.84} \\
\hline
\end{tabular}
\label{table:result-cicids-oneshot}
\end{table}

\subsection{Vehicle Network Data Packet Detection Experiment}

\begin{table}[h]
\centering
\caption{Car-hacking dataset split details}
\begin{tabular}{ccc}
\hline
Type & Training & Testing \\ 
\hline
R & 561466 & 140366 \\ 
DoS & 23601 & 5900 \\ 
Fuzzy & 19699 & 4925 \\ 
RPM & 26031 & 6508 \\ 
gear & 23955 & 5989 \\ 
\hline
\end{tabular}
\label{table:car-data-detail}
\end{table}

\textbf{Dataset:} This experiment used the car-hacking-dataset~\cite{song2020vehicle}, which was constructed by executing various typical message injection attacks in a real vehicle environment while simultaneously collecting CAN bus traffic. The attack scenarios cover the most common and representative attack types in the current vehicle network environment, including Denial of Service (DoS) attacks, fuzzy attacks, spoofing tachometer readings, and spoofing gear information. DoS attacks cause network congestion by continuously sending fake messages, fuzzy attacks disrupt communication using random CAN IDs and data fields, while spoofing attacks tamper with critical speed and gear information to mislead drivers. To ensure data quality, each attack scenario was repeated 300 times, lasting 3 to 5 seconds each time, while all data frames on the CAN bus were fully recorded at a high sampling frequency. Attack samples coexist with normal driving data, which can be used to evaluate the performance of anomaly detection algorithms. This dataset contains hundreds of thousands of messages spanning over 30 minutes, comprehensively reproducing the real situation of vehicles subjected to network attacks.

\textbf{Experimental Setup:} The main purpose of this experiment is to verify the transferability of our model, that is, whether it can be used directly in different scenarios without modifying the model itself. We found that by directly using data from new scenarios to train the model, cross-scenario application can be achieved. In the experiment, we still chose multi-classification tasks as the training objective of the model to evaluate its performance in vehicle network intrusion detection. We continued to use the parameters from Table \ref{table:train_setting}, and the specific dataset details are shown in Table \ref{table:car-data-detail}.

\textbf{Results Analysis:} The experimental results are shown in Table \ref{table:result-can}. Our model achieved 1.00 in Precision, Recall, and F1-score, performing comparably with other advanced methods such as transfer\_CNN~\cite{yang2022transfer} and D\_CNN~\cite{yang2022transfer}. This indicates that our model has excellent transferability and can achieve outstanding performance in vehicle network data packet detection tasks. It's worth noting that despite the differences in features and distribution between vehicle network data and network traffic data from previous experiments, our model could adapt to new scenarios and achieve excellent results without any modifications. This demonstrates the robustness and generalization ability of our model, which can effectively capture key patterns and features in different types of data, reflecting its universality. Moreover, our model does not require additional preprocessing operations, making it a completely end-to-end model, greatly simplifying the training process.

\begin{table}[h!]
\centering
\caption{CAN data packet detection experimental results}
\begin{tabular}{ccccc}
\hline
Method & Precision & Recall & F1-score \\
transfer\_CNN~\cite{yang2022transfer} & 1.00 & 1.00 & 1.00 \\
D\_CNN~\cite{yang2022transfer}  & 0.99 & 0.99 & 0.99 \\
\textbf{Ours} & \textbf{1.00} & \textbf{1.00} & \textbf{1.00} \\
\hline
\end{tabular}
\label{table:result-can}
\end{table}

\subsection{Interpretability Experiment}

\begin{figure}[h!]
    \centering
    \includegraphics[width=1\textwidth]{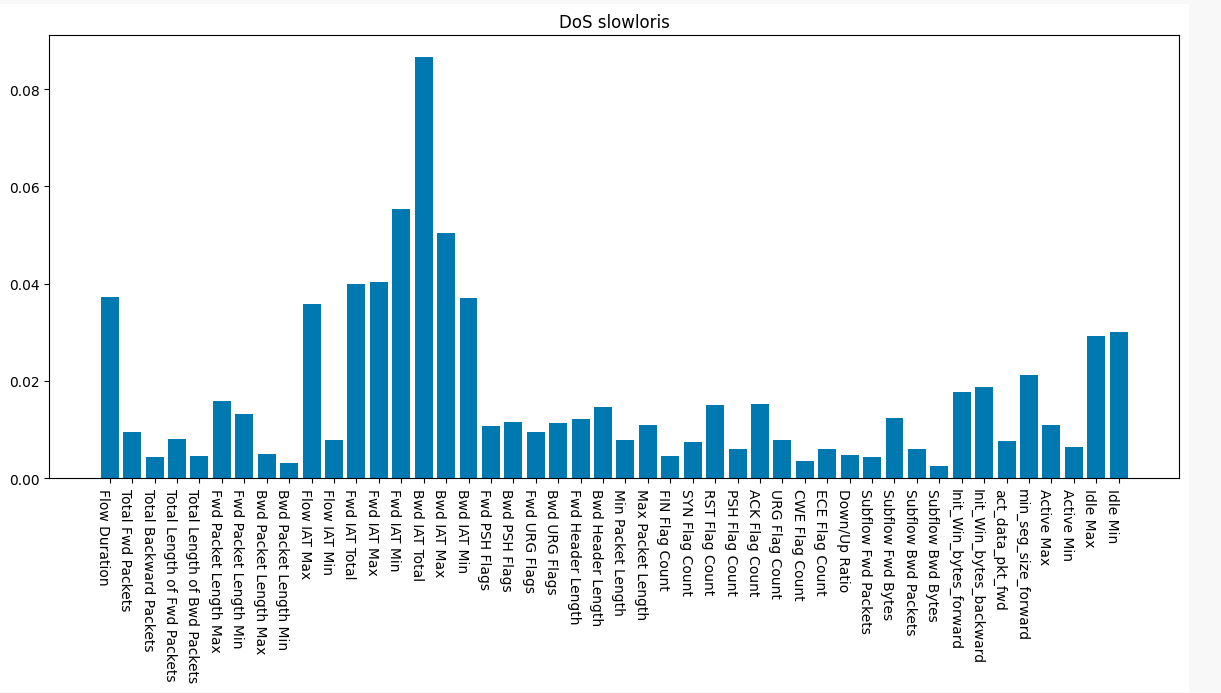}
    \caption{Network anomaly detection attention visualization}
    \label{fig:cicids_vis}
\end{figure}

\begin{figure}[h!]
    \centering
    \includegraphics[width=1\textwidth]{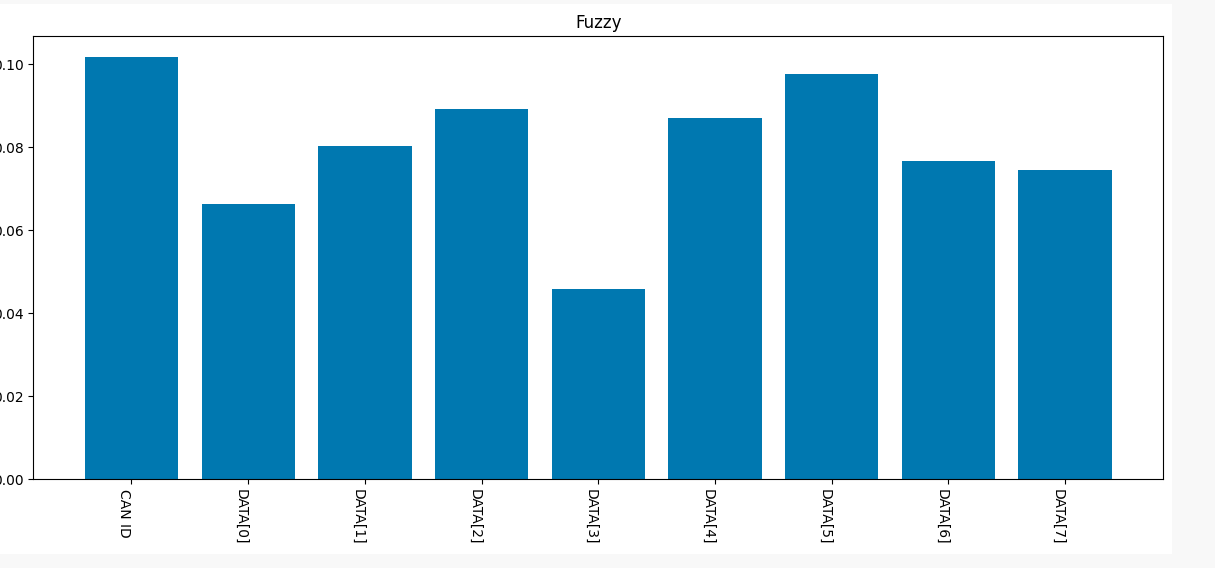}
    \caption{CAN detection attention visualization}
    \label{fig:can_vis}
\end{figure}

This experiment mainly aims to analyze the interpretability of the representations learned by the model. We extracted the weights of the trained models from experiments 1 and 2 and visualized the attention weights, as shown in Figures \ref{fig:cicids_vis} and \ref{fig:can_vis}. Through the visualization analysis of the model's attention weights in these two different scenarios, we can gain a deeper understanding of the key features that the model focuses on when performing intrusion detection. This not only helps to verify the effectiveness of the model but also provides important guidance for further model optimization.

In the analysis of the CICIDS-2017 dataset (Figure \ref{fig:cicids_vis}), we observed that when detecting DoS slowloris attacks, the model mainly focused on features related to packet arrival time (IAT). This focus distribution highly corresponds to the characteristics of slowloris attacks, as this type of attack usually manifests as abnormal packet arrival patterns. The model successfully captured this key feature, demonstrating its effectiveness in identifying such complex attacks.

In contrast, in the analysis of the vehicle network (CAN) dataset (Figure \ref{fig:can_vis}), the model showed balanced attention to multiple data fields. In particular, the CAN ID and certain specific data fields showed higher importance. This balanced attention distribution reflects the complexity of vehicle network communication while also indicating that the model can adapt to the characteristics of different network environments, considering multiple key factors when analyzing CAN bus data.

The results of these two sets of experiments reveal that our model has good adaptability and flexibility. It can automatically adjust its focus according to different network environments and potential attack types, thereby effectively identifying various possible anomalous behaviors. This adaptability is crucial for building a universal and effective intrusion detection system, especially when facing increasingly complex and diverse network threats.

Through these visualization analyses, we not only verified the effectiveness of the model in different scenarios but also gained valuable insights, pointing the direction for further model optimization. For example, we can consider selectively enhancing the weights of certain key features or introducing new features to capture more complex attack patterns. At the same time, these findings also provide important references for designing more precise and targeted defense strategies.

\section{Conclusion}

The NIDS-GPT model has demonstrated significant potential and advantages in the field of network intrusion detection. Through innovative architecture design and training strategies, this model can not only effectively handle challenging tasks such as imbalanced data and one-shot learning but also shows good scalability and transfer learning capabilities. The experimental results on the CICIDS-2017 and OTIDS datasets prove the effectiveness of the model, especially in maintaining a high level of detection accuracy even in extremely imbalanced data situations. Furthermore, the exploration of model interpretability provides a foundation for improving its credibility and operability in practical applications.

However, future research still needs to further explore the model's performance in more complex network environments, as well as how to further improve the model's efficiency and real-time performance. Overall, NIDS-GPT provides a promising new approach for the field of network security and is expected to play an important role in future practical applications.

\bibliographystyle{unsrt}   
\bibliography{main}   

\end{document}